\documentclass{sn-jnl}

\usepackage{natbib}
\usepackage{booktabs}

\usepackage{graphicx}%
\usepackage{multirow}%
\usepackage{amsmath,amssymb,amsfonts}%
\usepackage{amsthm}%
\usepackage{mathrsfs}%
\usepackage[title]{appendix}%
\usepackage{xcolor}%
\usepackage{textcomp}%
\usepackage{manyfoot}%
\usepackage{multicol}
\usepackage{algorithm}%
\usepackage{algorithmicx}%
\usepackage{algpseudocode}%
\usepackage{listings}%
\usepackage{float}
\usepackage{subfig}

\begin{document}

\title{
Advances in Kth nearest-neighbour clutter removal
}


\author{\fnm{Nicoletta} \sur{D'Angelo}}


\affil{\orgdiv{Department of Economics, Business and Statistics}, \orgname{University of Palermo}, \orgaddress{\city{Palermo},  \country{Italy}}}


\abstract{We consider the problem of feature detection in the presence of clutter in spatial point processes. 
Classification methods have been developed in previous studies. Among these, \cite{Byers1998} models the observed Kth nearest neighbour distances as a mixture distribution and classifies the \textit{clutter} and \textit{feature} points consequently. 
In this paper, we enhance such approach in two manners. First, we propose an automatic procedure for selecting the number of nearest neighbours to consider in the classification method by means of segmented regression models. 
Secondly, with the aim of applying the procedure multiple times to get a ``better" end result, we propose a stopping criterion that minimizes the overall entropy measure of cluster separation between clutter and feature points.
The proposed procedures are suitable for a feature with clutter as two superimposed Poisson processes on any space, including linear networks. We present simulations and two case studies of environmental data to illustrate the method. 
}

\keywords{Changepoints; Clutter; Entropy measure; Feature; Spatial point processes}



\maketitle

\section{Introduction}

Point processes are defined as random collections of points within a measurable space. They have found widespread utility in describing a diverse range of naturally occurring phenomena across various fields. These applications include epidemiology, ecology, forestry, mining, hydrology, astronomy, and meteorology, among others \citep{cox1980point,ripley2005spatial,daley2003introduction,moller2003statistical,schoenberg2008description,tranbarger2010computation}. 

In spatial point processes, each point denotes the location of a specific object or event, such as a tree or a sighting of a species \citep{ripley2005spatial,cressie2015statistics,diggle1976statistical}.

The aim is typically to learn about the mechanism that generates these events \citep{moller2003statistical,diggle1976statistical,illian:penttinen:stoyan:stoyan:08}.
The first step is usually to learn about its first-order characteristics, studying the relationship of the points with the underlying environmental variables that describe the observed heterogeneity. 
When the purpose of the analysis is to describe the possible interaction among points, that is, if the given data exhibit spatial inhibition or aggregation, the second-order properties of the process are analysed.

One of the main interests of spatial point pattern analysis is identifying features surrounded by clutter. 
The conventional terminology is that a \textit{feature} is a point of the pattern or process of interest, and \textit{clutter} (also called \textit{noise}) consists of extraneous points that are not proper to the pattern of interest. For instance, detecting surface minefields from an image from a reconnaissance aircraft can be processed to obtain a list of objects, some of which may be mines and others any other type of object \citep{1997-AF, Byers1998}.

For spatial point processes, this problem has been addressed differently, either denoted by \textit{feature detection} or \textit{clutter removal}.  \cite{1997-AF} developed a method to find the maximum likelihood solution using Voronoi polygons. \cite{dasgupta1998detecting} used model-based clustering to extend the methodology proposed by \cite{banfield1993model}.  While these methods are based on some limiting assumptions, \cite{Byers1998} adopted a different approach in which they estimated and removed the clutter without making any assumptions about the shape or number of features. More recently, \cite{gonzalez2021classification} considers the local contributions of the pair correlation function as functional data and describes two classification procedures to separate features from clutter points.

Among these,  \cite{Byers1998}'s approach represents a simple and intuitive method for estimating regions of different point densities in a point process, with the very useful feature being potentially for easy use in higher dimensions.
Their solution uses $K$th nearest-neighbour distances of points in the process to classify them as clutter or otherwise. Such distances are modelled as a mixture distribution, the parameters of which are estimated by a simple EM algorithm.
However, as pointed out by the authors, the value of $K$ to be used must be specified by the user, and though they gave some guidelines, this area could benefit from further investigation.
Moreover, they highlight another extension which shows promise, that is, the possibility of applying the procedure multiple times to get "better" end results. This would treat the estimated feature as a new dataset and apply the same method to this. 

Given the above, this paper aims at enhancing the approach of \cite{Byers1998} in two ways.
First, we propose a procedure to automatically select the number of nearest neighbours $K$ to consider in the classification algorithm by means of segmented regression models. Secondly, we consider the further extension of applying the procedure multiple times. 
In this context, a stopping criterion is needed, and we propose such a criterion based on an entropy measure of cluster separation.

All the analyses are carried out through the statistical software \cite{R} and are available from the author.

The structure of the paper is as follows. Section \ref{sec:preliminaries} presents the preliminaries, including \cite{Byers1998}'s method for feature detection and the basics about segmented regression models. Section \ref{sec:proposal} introduces the proposed methodologies: the selection of the nearest neighbour to consider, trough segmented regression, and the stopping criterion to apply when the procedure is run iteratively. Section \ref{sec:simulations} shows a simulation study, and Section \ref{sec:application} shows two case studies on environmental data. Finally, Section \ref{sec:conclusions} presents the conclusions. 

\section{Preliminaries}\label{sec:preliminaries}

\subsection{Kth nearest neighbour clutter removal}

Let $u$ be a point location in the two-dimensional plane 
and  $D_K$ be the distance of its $K$th nearest neighbour. If $D_K$ is greater than the spatial range $r_u$, then, there must be one of $0, 1, \ldots, K - 1$ points at a distance less than $r_u$. For all $u \in W$, with $W$ being the spatial window, and $x \in [0,\infty)$, the $K$th nearest neighbour distribution approximation is given by
\begin{equation*}
\mathbb{P}(D_K\geq x)=\sum_{k=0}^{K-1} \frac{e^{-\lambda  \pi  x^2}(\lambda \pi x^2)^k}{k!}=1-F_{D_K}(x),
\label{eq:cumulatedensity}
\end{equation*}
where $\mathbb{P}(D_K\geq x)$ is the probability that the $K$th nearest neighbour point falls out of the disk $b(u,x)$ with $|b(u,x)|=x$, assuming that this disk exists around $u$. If the $K$th nearest neighbour point of $u$ is outside $b(u,x)$, it is also outside $b(u,r_u)$. 

Accordingly, the density $f_{D_K}(x)$ can be found as
\begin{align}
f_{D_K}(x)&=\frac{e^{-\lambda \pi x^2}2(\lambda \pi)^K x^{2K-1}}{(K -1 )!},
\label{eq:density}
\end{align}
and therefore $Y\sim \Gamma(K, \lambda \pi)$, with $Y = (D_K)^2$. Having a closed-form and the Gamma distribution properties, the maximum likelihood estimation of the rate given the observed values of $D_K$ is straightforward. Indeed, the maximum likelihood estimate of $\lambda$ is
\begin{equation*}
    \hat{\lambda}=\frac{nK}{\pi \sum^n_{i=1}d^2_i},
\end{equation*}
where $d_i$ is the $i$th observed $K$th nearest neighbour distance.

We assume two types of processes to be classified through a mixture of the corresponding $K$th nearest neighbour distances coming from the clutter and feature, which are two superimposed Poisson processes. Therefore, based on equation \eqref{eq:density}, we assume that
\begin{equation*}\label{mix-dist}
D_K \sim p \Gamma^{1/2}(K, \lambda_1 \pi)+(1-p) \Gamma^{1/2}(K, \lambda_2 \pi),
\end{equation*}
where $\lambda_1$ and $\lambda_2$ are the intensities of the two homogeneous Poisson point processes (\textit{clutter} and \textit{feature}) and $p$ is the constant that characterizes the postulated distribution of the $D_K$.

\begin{figure}[H]
				\centering
\includegraphics[width=.425\textwidth]{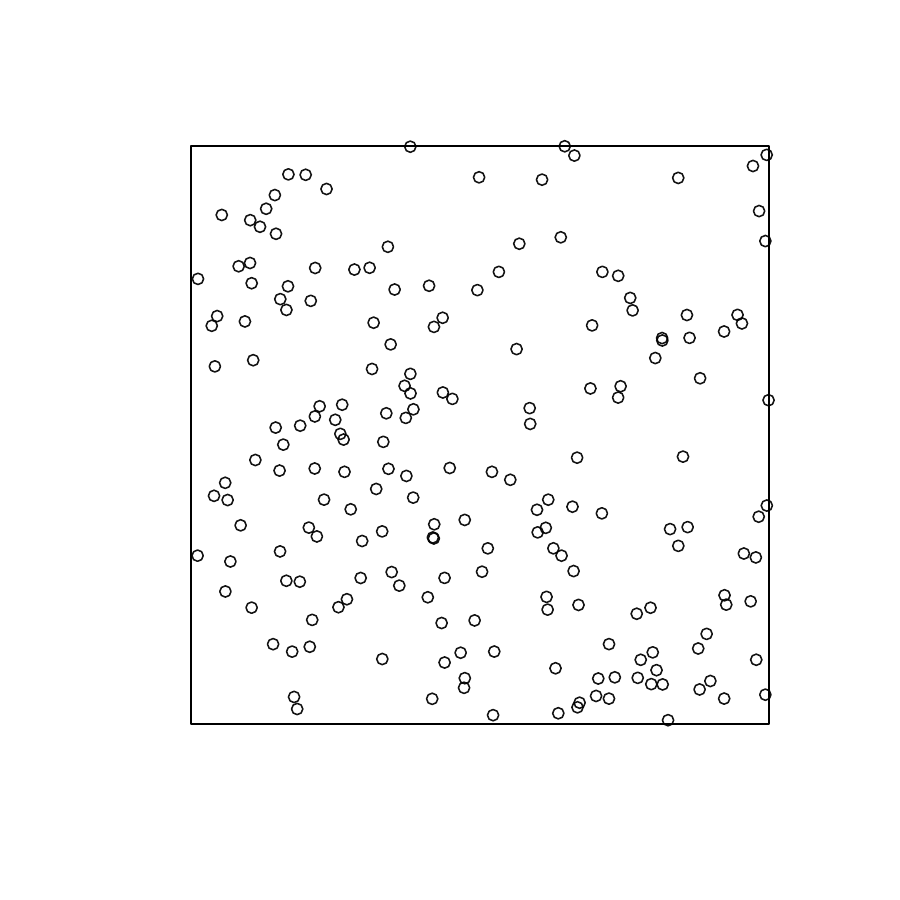}
\includegraphics[width=.49\textwidth]{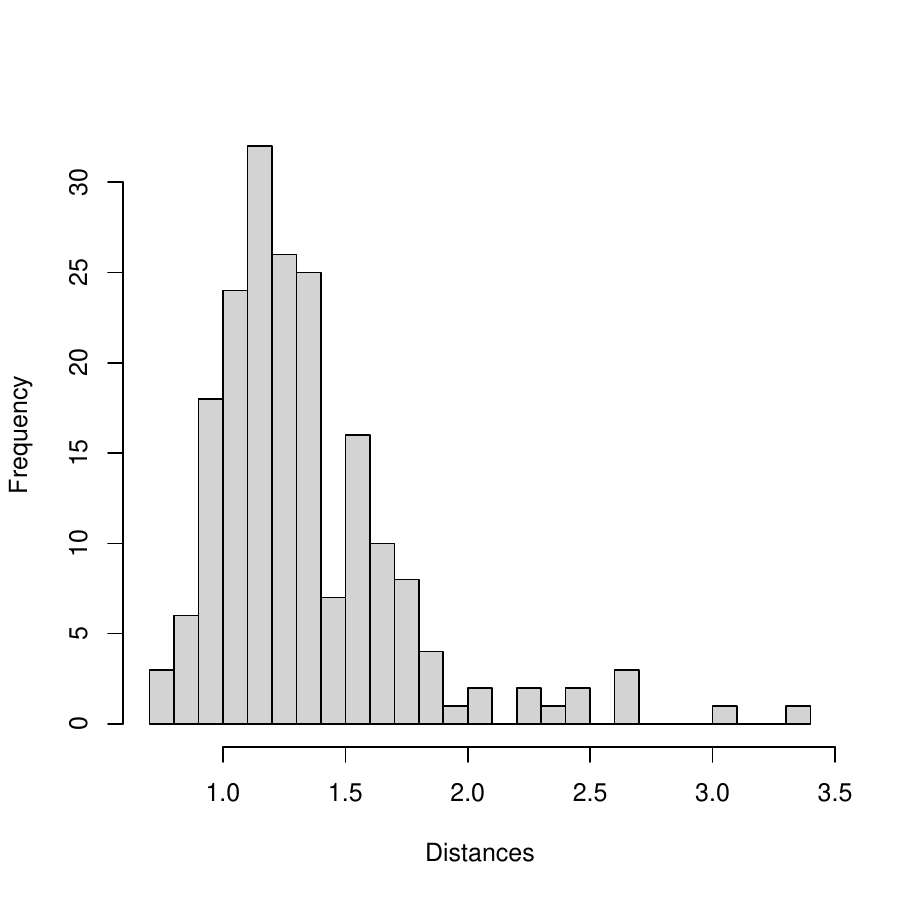}\\
\includegraphics[width=.49\textwidth]{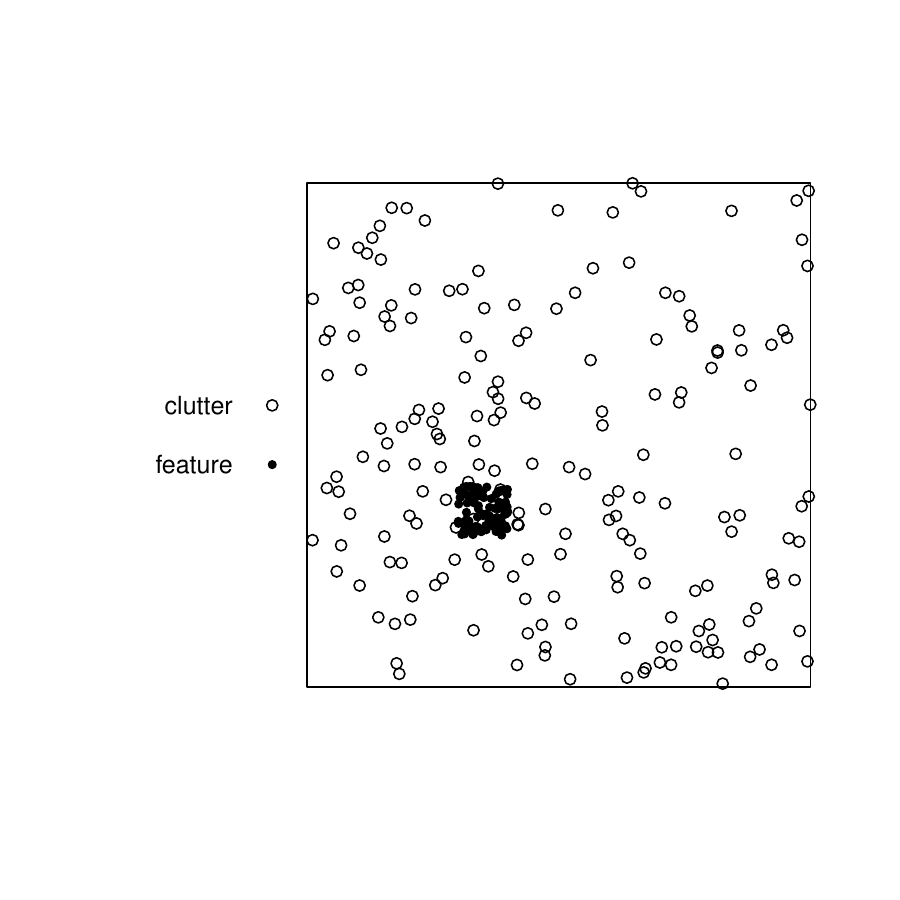}
\includegraphics[width=.49\textwidth]{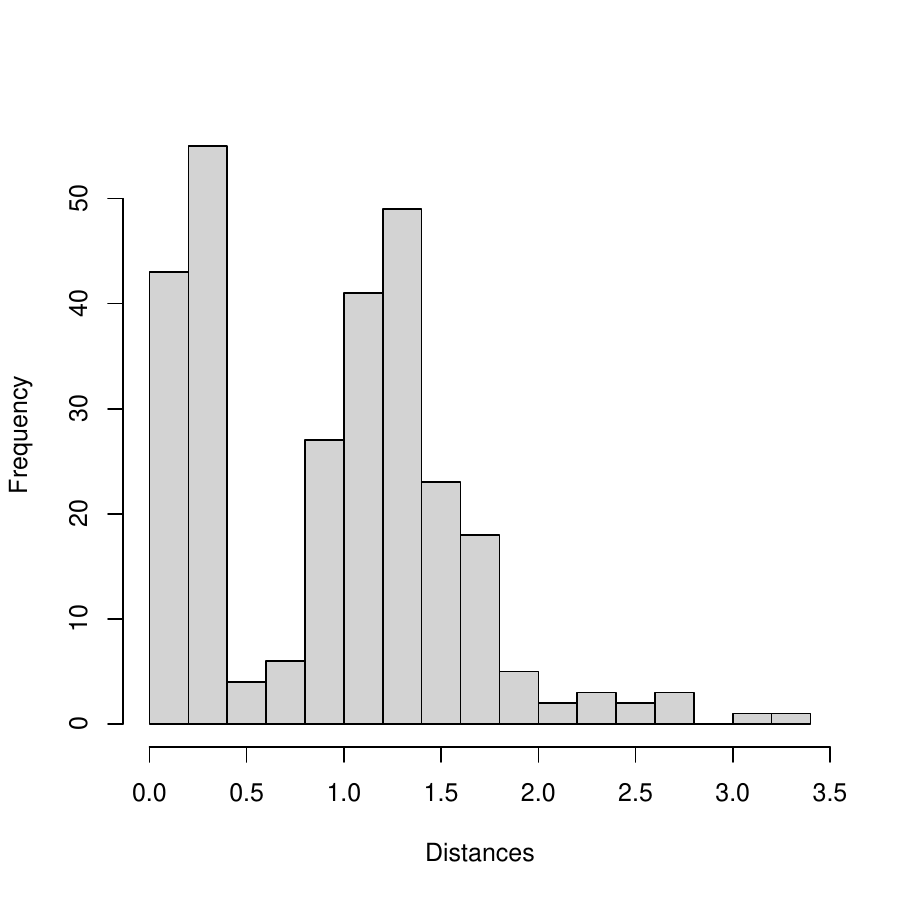}
				\caption{\textit{Top panels}: Simulated homogeneous Poisson process and its distances from the 10th nearest neighbour; \textit{Bottom panels}: Simulated clutter Poisson process with a feature Poisson pattern superimposed their distances from the 10th nearest neighbour.}
				\label{fig:clutter}
			\end{figure}

A graphical example is given in Figures \ref{fig:clutter}. 
In particular, the top panels of Figure \ref{fig:clutter} display a simulated homogeneous Poisson process, with 200 expected points and the distances among all the points of the pattern and their 10th nearest neighbours. The histogram of the distances shows an unimodality around the value $1.5$.

Then, the bottom panels of Figure \ref{fig:clutter} show what we assume in equation \eqref{mix-dist}, that is, a pattern that is obtained by the superposition of the previously simulated Poisson process on the $[0,10]\times[0,10]$ square (what we shall call \textit{clutter}), and another Poisson process (what we shall call \textit{feature}), with 100 expected number of points, on the unit square. As expected, the computed distances show an evident bimodality, ascribable to the different distances among points of the clutter and of the features with their 10th nearest neighbours. The underlying assumption is that the new modality around the value 0.25 is attributable to the points of the \textit{feature}.

The parameters $\lambda_1, \lambda_2$ and $p$ associated with the mixture are estimated using an EM algorithm \citep{DLR1977}, wherein we use the closed-form of a Gamma distribution in the expectation step. On the other hand, let $\delta_i \in \{0, 1\}$ be the two classification components for each data point, where $\delta_i= 1$ if the $i$th point belongs to the feature and $\delta_i= 0$ otherwise. Thus, each data point has an observation $d_i$ of $D_K$ and an unknown $\delta_i$. Hence, the $\mathbb{E}$ step of the algorithm consists of
\begin{equation*}
\mathbb{E}[\hat \delta_i^{(t+1)}]=\frac{\hat p^{(t)}f_{D_K}(d_i;\hat \lambda_1^{(t)})}{\hat p^{(t)}f_{D_K}(d_i;\hat \lambda_1^{(t)})+(1-\hat p^{(t)})f_{D_K}(d_i;\hat \lambda_2^{(t)})},
\end{equation*}
and the maximization $M$ step consists of
\begin{equation*}
\hat \lambda_1^{(t+1)}=\frac{K \sum_{i=1}^n\hat \delta_i^{(t+1)}}{\pi \sum_{i=1}^n d^2_i\hat \delta_i^{(t+1)}}, \quad  \hat\lambda_2^{(t+1)}=\frac{K \sum_{i=1}^n(1-\hat \delta_i^{(t+1)})}{\pi \sum_{i=1}^n d^2_i(1-\hat \delta_i^{(t+1)})} \quad 
\end{equation*}
and
\begin{equation*}
\quad \hat p^{(t+1)}=\frac{\sum_{i=1}^n\hat \delta_i^{(t+1)}}{n}.
\end{equation*}
An intuitive classification test criterion would classify the points according to the mixture component where the distances have the highest densities. We are mainly interested in identifying the feature points in this proposed classification approach; consequently, we do not consider edge effects because feature points, in practice, are predominantly far from the edges. Additionally, for large $n$, the convergence of the EM algorithm is good since it takes less time to arrive at an approximately acceptable solution, also with the fewest number of iterations.

The following steps implement the classification procedure:
\begin{enumerate}
    \item \label{iter:1} Choose a value of $K$.
    \item \label{iter:2} Compute the $K$th nearest-neighbour distances for each point in the point pattern.
    \item \label{iter:3} Apply the EM algorithm for estimating $\lambda_1$, $\lambda_2$, and $p$.
    \item \label{iter:4} Classify the points according to whether they have a higher density under the feature or clutter component of the mixture.
    \item \label{iter:5} Repeat the steps \ref{iter:1}-\ref{iter:4} iteratively as desired. 
\end{enumerate}

\subsection{Segmented regression models}\label{sec:segmented}

Segmented, or broken-line models, are regression models where the relationships between the response and one or more explanatory variables are piecewise linear and, as such, represented by two or more straight lines connected at unknown points. These models are a common tool in many fields, including epidemiology, occupational medicine, toxicology and ecology, where usually it is of interest to assess threshold values where the effect of the covariate changes.
The main advantage of this approach is the easy interpretation given by two components, i.e. changepoints and slopes.

The segmented linear regression is expressed as
\begin{equation}
\label{eq:seg.glm}
g(E[Y|x_i,z_i])= \alpha + z_i^{T}\theta+\beta x_i+\sum_{m=1}^{M_0}\delta_m (x_{i,m}-\psi_m)_+  
\end{equation}
where $g$ is the link function,  $x_i$ is the broken-line covariate, and $z_i$ is a covariate vector whose relationship with the response variable is a non-broken-line. We denote by  $M_0$ the true number of changepoints and by $\psi_m$  the $M_0$ locations of the changepoints in the observed phenomenon. 
These $M_0$ are selected among all the possible values in the range of $x$.
The term $(x_i-{\psi}_m)_+$ is defined as $\sum_i I(x_i>{\psi}_m)$ that is $(x_i-{\psi}_m)I(x_i>{\psi}_m)$.
The parameter estimates $\boldsymbol{\theta}$ represent the non broken-line effects of $z_i$,
 $\beta$ represents the effect for $x_i < \psi_1$, while $\boldsymbol{\delta}$ is the vector of the differences in the effects.

The parameters to be estimated usually are: the number of changepoints $M_0$; their locations $\psi_m$; and the broken-line effects, represented by $\beta$ and $\boldsymbol{\delta}$. For the estimation procedure, we refer to \cite{muggeo2003estimating}.

In this paper, we focus on the sole objective of estimating the location of a unique changepoint, that is, $\psi_m$, with $M_0$ fixed at $1$, and no further covariates $z_i$.

\section{Proposed approaches}\label{sec:proposal}

This section is devoted to the enhancements of the EM algorithm for the classification of \textit{clutter} and \textit{feature}. 

Section \ref{sec:proposal1} solves the problem of Step \ref{iter:1} of the algorithm by suggesting an approach to select $K$ automatically.

Section \ref{sec:proposal2} illustrates a stopping criterion to solve the iterative problem of Step \ref{iter:5}.
By means of the entropy measure of cluster separation employed in Section \ref{sec:proposal1}, we provide a simple and intuitive way to state that the current iteration is enough to separate clutter and feature correctly.

\subsection{Selecting K through changepoint detection}\label{sec:proposal1}

The development of the method of \cite{Byers1998} assumes a proper value of $K$ priorly chosen. The natural way to choose the suitable $K$th neighbour is by analysing several increasing values of $K$ and then selecting the $K$ after which no improvement is found.

\begin{figure}[H]
\centering
    \includegraphics[width=.4\textwidth]{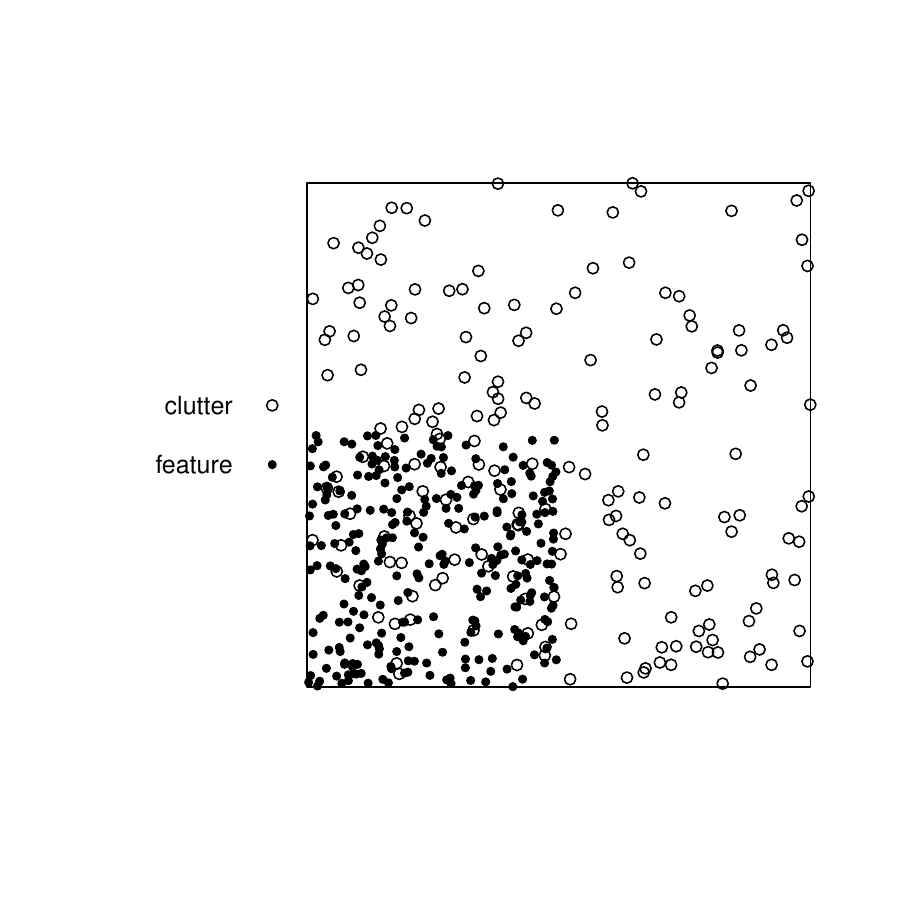}
\includegraphics[width=.575\textwidth]{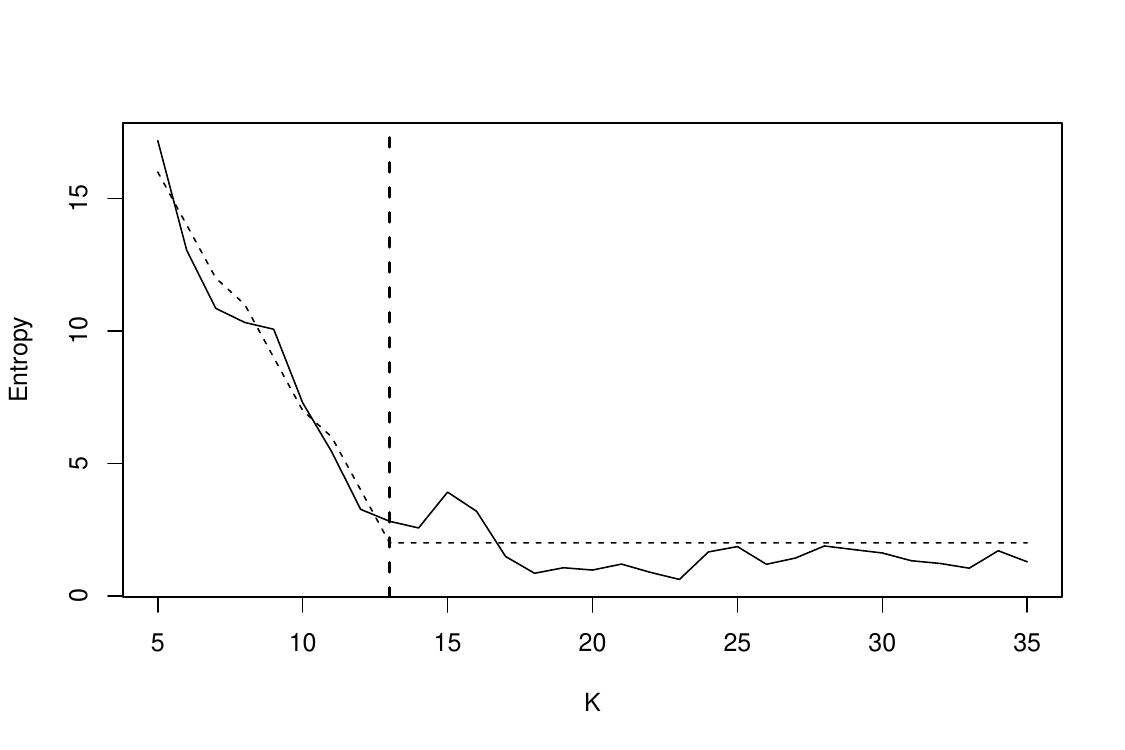}
\caption{\label{fig:seg} \textit{Left panel}: Simulated clutter Poisson process with a feature Poisson pattern superimposed; \textit{Right panel}: Entropy values of the simulated pattern. The black line represents the observed entropies, and the dotted line represents the estimated segmented model. The vertical line indicates the estimated changepoint of $\hat K=13$.}
\end{figure}

In the literature, there are several methodological proposals for this target; in this work, we use an entropy-type measure of separation introduced in \cite{celeux1996entropy} given by  
\begin{equation*}
    S = - \sum_{i=1}^{n} \delta_i \log_{2}(\delta_i),
    \label{eq:entropy}
\end{equation*}
where $\delta_i$ are the probabilities of being in the first component of the mixture in equation \eqref{mix-dist}, which is the feature. As stated by \cite{Byers1998}, plotting the entropies sequentially and looking for a levelling-off changepoint in the graph is an easy way to choose $K$.
 An example of this procedure is shown in Figure~\ref{fig:seg}, where the classification entropies for values of $K$ up to $35$ are plotted (\textit{right panel}) for a simulated point pattern (\textit{left panel}).

However, such a graphical assessment is not formalized and, therefore, not generalizable and reproducible.

Therefore, our first proposal consists of the optimal $K$ being estimated by fitting a segmented regression model as 
\begin{equation*}
    \mathbb{E}\left[Y|x_i \right]=\beta+\delta(x_i-{\psi})I(x_i>{\psi}),
\end{equation*}
where the interest is estimating a unique changepoint $\psi$,
after which the slope $\beta + \delta$ is constrained to be equal to zero. As depicted in Figure~\ref{fig:seg}, the observed response variable is the entropy level, modelled as a function of the number of nearest neighbours. We implemented this automatic option using the function \texttt{segmented} of the package \texttt{segmented} \citep{muggeo2008segmented}. In this case, the fitting of the segmented model leads to a $\hat{K} = 13$.

\subsection{Stopping criterion for the iterative procedure}\label{sec:proposal2}

Let's consider again the simulated pattern of Figure \ref{fig:seg} (a).

We run the EM algorithm iteratively to see if we can get better results compared to running the algorithm only once.

\begin{figure}[H]
\centering
\includegraphics[width=\textwidth]{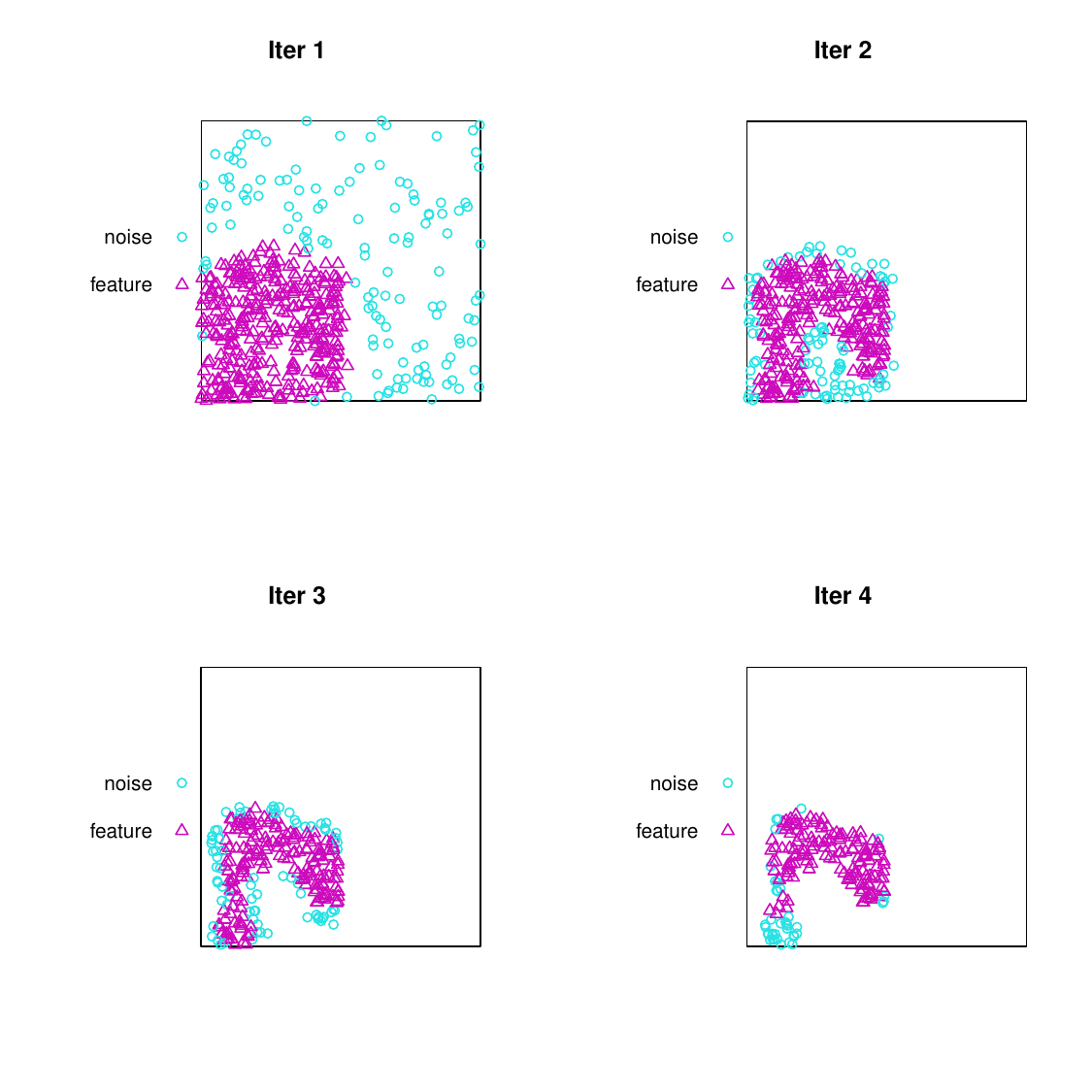}
\caption{\label{fig:stopping0} Points of the simulated pattern classified through the EM algorithm up to four iterations. Blue denotes the \textit{clutter}/\textit{noise} points, and pink denotes the \textit{feature} points.}
\end{figure}

Figure \ref{fig:stopping0} shows the output of the EM procedure run iteratively up to 4 times.

Note that we also let the algorithm automatically select $K$ at each step, and the set of estimated nearest neighbours at each iteration is equal to $\hat{K} = \{13, 19 ,24 ,34\}$.
As evident from Figure \ref{fig:stopping0}, the first iteration looks sufficient to spot the majority of the true feature points.
To corroborate this statement, Table \ref{tab:acc} contains the true-positive rate (TPR), false-positive rate (FPR), and accuracy (ACC), respectively defined as
\begin{equation*}
    TPR = \frac{\text{true positives}}{\text{positives}}, \quad FPR = \frac{\text{false negatives}}{\text{negatives}}, \quad ACC = \frac{\text{true positives and negatives}}{\text{positives and negatives}}.
\end{equation*}
We, of course, wish to have TPR and ACC close to 1 and FPR close to 0. 

\begin{table}[h]
\centering
\caption{True-positive rate (TPR), false-positive rate (FPR), and accuracy (ACC) resulting from the application of the EM algorithm iteratively to the simulated point pattern of Figure \ref{fig:seg}.}
\begin{tabular}{c|ccc}
\toprule
Iteration&TPR & FPR & ACC\\
\midrule
1&0.982 & 0.349 & 0.849 \\
2&0.746 & 0.240 & 0.752 \\
3&0.539 & 0.146 & 0.666 \\
4&0.415 & 0.125 & 0.601 \\
\bottomrule
\end{tabular}
\label{tab:acc}
\end{table}

These results, of course, confirm that one iteration is sufficient to classify points into clutter and features correctly.

However, in real-life applications, such classification rates cannot be computed.

Therefore, our proposed stopping criterion to automatically select the number of iterations to run is formalized as follows.

Consider a measure of the overall entropy of a unique iteration.
Let's denote by $K_{set}$ the set of possible $K$ values investigated.
Then, for the $j$th iteration, regardless of the $K$ having had to be estimated or fixed, we compute the Entropy measure in equation \eqref{eq:entropy} for each $K \in K_{set}$. We denote the entropy measure obtained considering the $K$th nearest neighbour by $S_K$.
Then, the \textit{overall measure of entropy} $S_J$ of the $j$th iteration is just given by the sum of all the entropies computed for the set of $K$ values, namely
\begin{equation*}
    S_J = \sum_{K_{set}} S_K.
\end{equation*}
Note that $K_{set}$ is not indexed by $J$ as we assume the same set for each iteration.
The EM algorithm then stops at iteration $J$ whenever $S_{J + 1} > S_J$, that is, whenever the overall measure of the entropy of the next iteration exceeds the current one.

\begin{figure}[H]
\centering
\includegraphics[width=\textwidth]{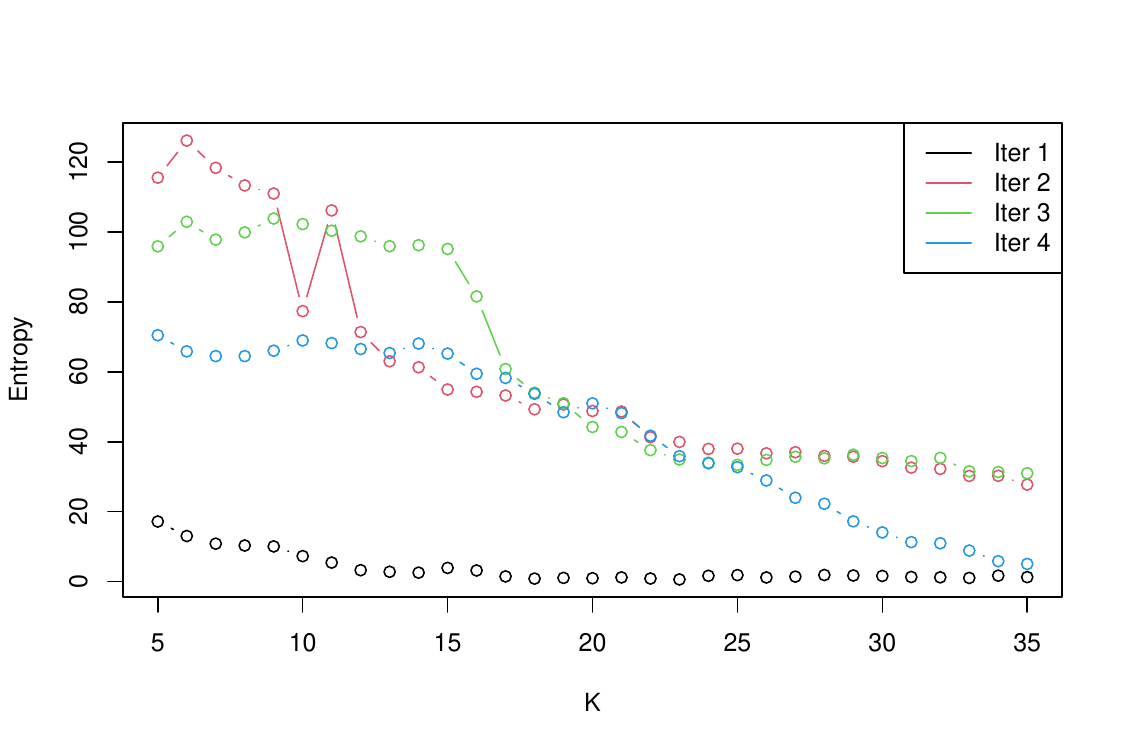}
\caption{\label{fig:stopping} Entropy values for all the investigated iterations.}
\end{figure}

Figure \ref{fig:stopping} gives a graphical representation and justification of the idea underlying this criterion.

The Figure shows that the first iteration provides the lower overall entropy values.
The results in Table \ref{tab:acc2} confirm this result, showing that $S_2 > S_1$.

\begin{table}[h]
\centering
\caption{Overall measures of entropy $S_{J}$ for the 4 iterations.}
\begin{tabular}{c|c}
\toprule
Iteration&$S_{J}$ \\
\midrule
\textbf{1}&\textbf{115}\\
2&1813 \\
3&1904 \\
4&1345 \\
\bottomrule
\end{tabular}
\label{tab:acc2}
\end{table}

Basically, the algorithm stops at the first iteration ($\hat{J} = 1)$ because it is the one providing the first value of total entropy $S_J$ that does not decrease at the following iteration.

Consider now another example where the clutter points are simulated in the unit square, and the feature points are simulated in a $[0.25,0.5]\times[0.25,0.5]$ window with a different intensity.

\begin{figure}[H]
\centering
\includegraphics[width=\textwidth]{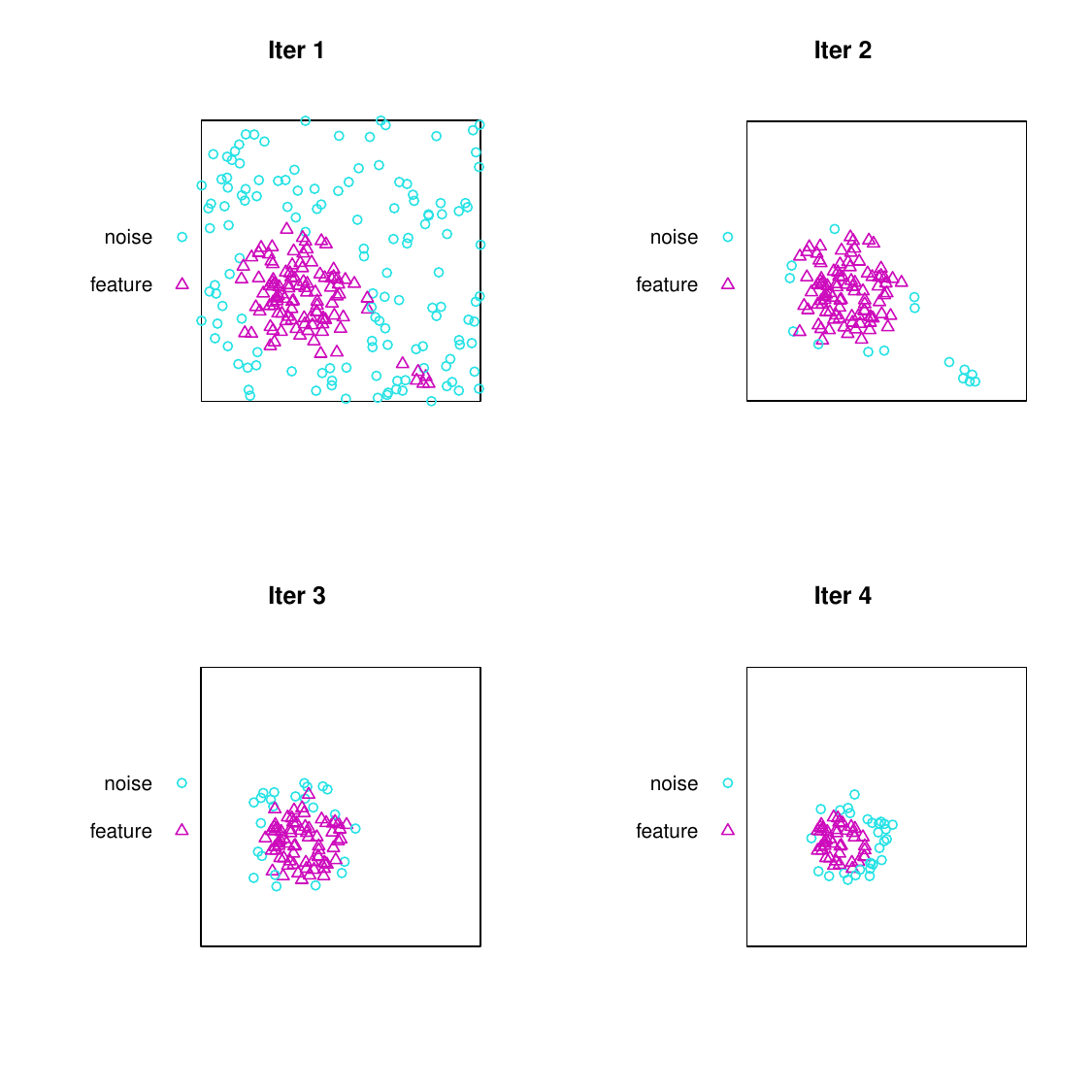}
\caption{\label{fig:stopping1} Points of the simulated pattern classified through the EM algorithm up to four iterations. Blue denotes the \textit{clutter}/\textit{noise} points, and pink denotes the \textit{feature} points.}
\end{figure}

Figure \ref{fig:stopping1} shows the points of such simulated pattern classified through the EM algorithm up to four iterations. 

Knowing the sub-window where the feature points have been simulated, we expect the stopping criterion to select the second iteration as the final one, as in the first iteration also points outside of the $[0.25,0.5]\times[0.25,0.5]$ window are classified as features.

\begin{figure}[H]
\centering
\includegraphics[width=\textwidth]{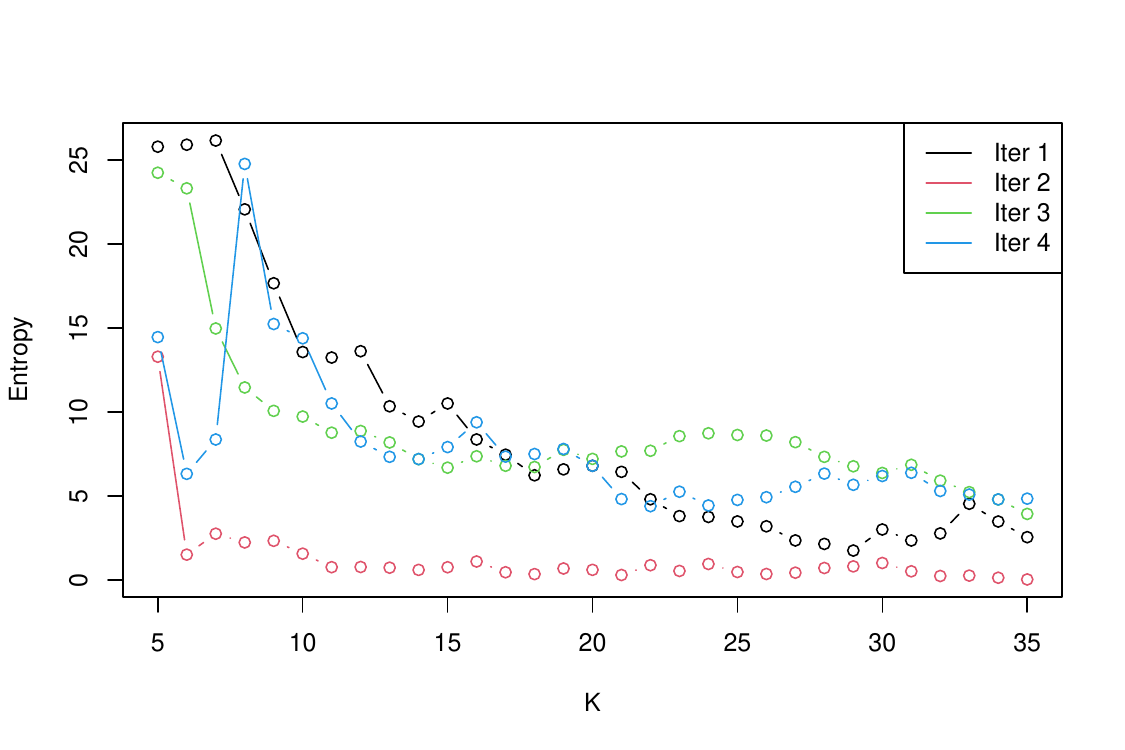}
\caption{\label{fig:stopping2} Entropy values for all the investigated interations.}
\end{figure}

Indeed, Figure \ref{fig:stopping2} and Table \ref{tab:acc3} confirm such expectation, indicating the second iteration as the one providing the value of $S_J$ after which the entropy tends to increase again. In other words, $S_2 < S_1$, but $S_3 > S_2$, therefore $\hat{J} = 2$. 

\begin{table}[h]
\centering
\caption{Overall measures of entropy $S_{J}$ for the 4 iterations.}
\begin{tabular}{c|c}
\toprule
Iteration&$S_{J}$ \\
\midrule
1&274\\
\textbf{2}&\textbf{38} \\
3&274 \\
4&242 \\
\bottomrule
\end{tabular}
\label{tab:acc3}
\end{table}

\section{Simulation study}\label{sec:simulations}

This section aims to study the proposed method's performance in terms of classification rates, considering different scenarios concerning both the generating processes and the ratio between the number of clutter and feature points generated. 
To this end, we simulate under different such scenarios to obtain a comprehensive understanding of the results of the proposed algorithm in different settings.

The simulation setup is as follows.
 We simulate 200 patterns from clutter Poisson point processes with $\mathbb{E}[n_c]$ expected number of points.

 The feature point patterns, with  $\mathbb{E}[n_f]$  expected number of points, are simulated from the following processes:
\begin{enumerate}
    \item\label{scen:1} Poisson cluster process with $\kappa = 7.5 $ intensity of the Poisson process of cluster centres in the window $W_c =[0,1]$. Each cluster consists of $u = 20$ points in a disc of radius 0.2; 
    \item\label{scen:2} Poisson cluster process with $\kappa  = 15$ intensity of the Poisson process of cluster centre in the window $W_c =[0,1]$. Each cluster consists of $u = 10$ points in a disc of radius 0.2;
   \item\label{scen:3} Poisson processes in the sub-window $W_c =[0,0.5]$ with 150 expected number of points;
    \item\label{scen:4} Poisson processes in the sub-window $W_c =[0.25,0.5]$ with 20 expected number of points.
\end{enumerate}

\begin{figure}[H]
\centering
\vspace{-1cm}
\subfloat[Scenario \ref{scen:1}]{\includegraphics[width=.4\textwidth]{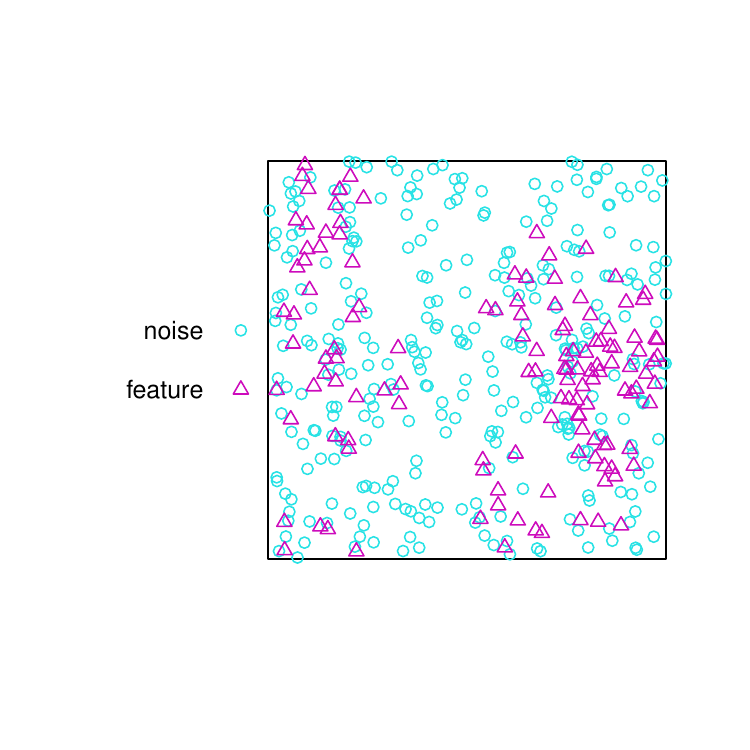}}
\subfloat[Scenario \ref{scen:2}]{\includegraphics[width=.4\textwidth]{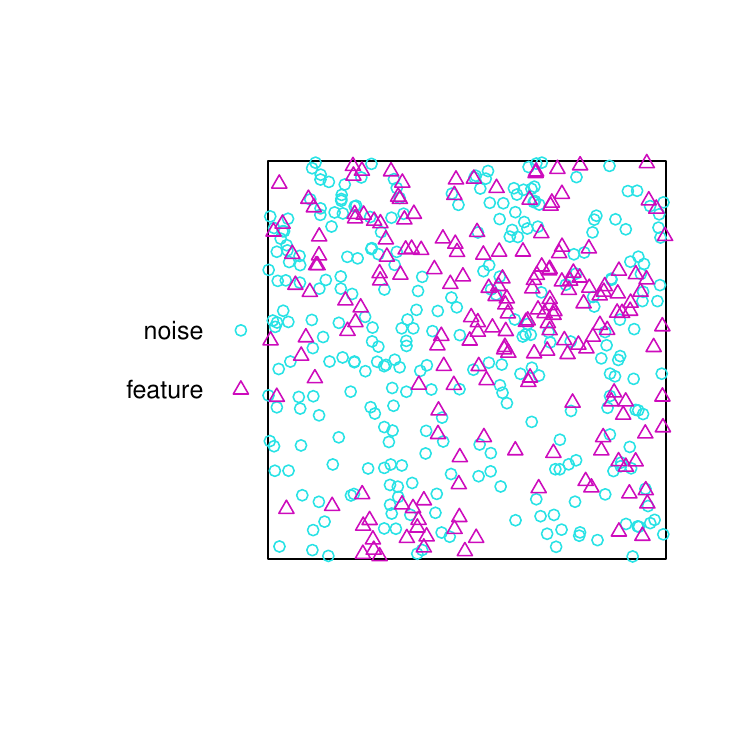}}\\
\subfloat[Scenario \ref{scen:3}]{\includegraphics[width=.4\textwidth]{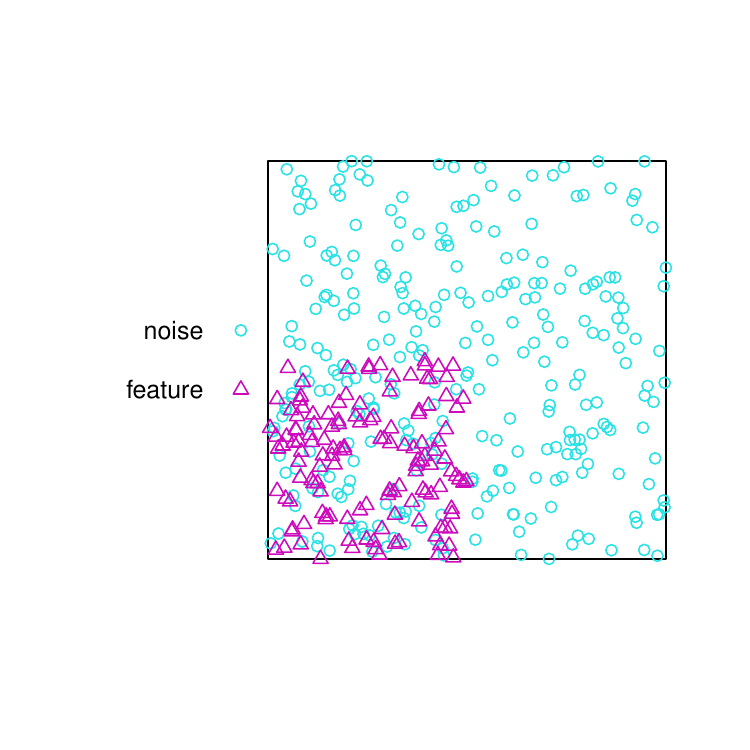}}
\subfloat[Scenario \ref{scen:4}]{\includegraphics[width=.4\textwidth]{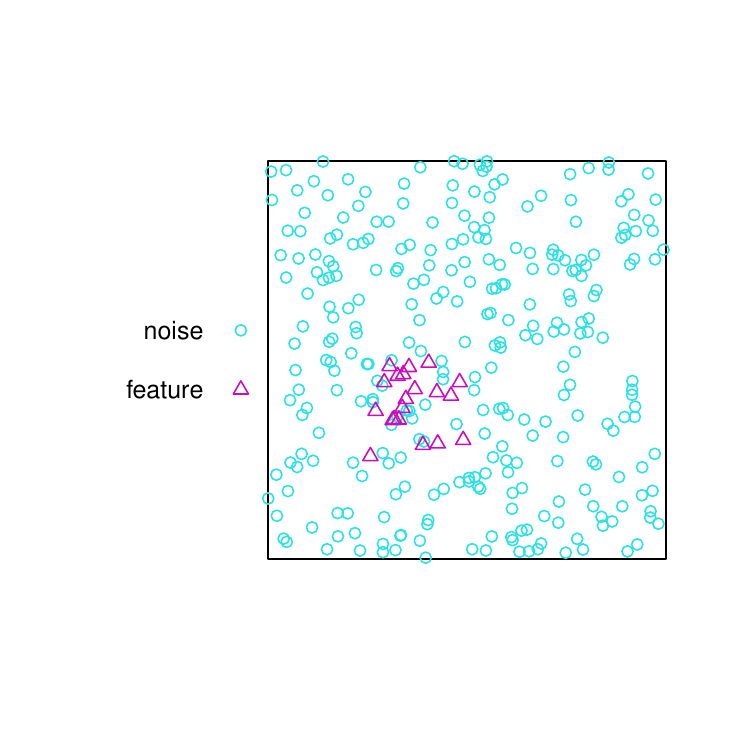}}
	\caption{Patterns simulated from the considered processes. Blue denotes the \textit{clutter}/\textit{noise} points, and pink denotes the \textit{feature} points.}
	\label{fig:simula}
\end{figure}
Examples of the simulated patterns are depicted in Figure 	\ref{fig:simula}.

We show the results of the proposed procedure in Table \ref{tab:1}, in terms of true-positive rate (TPR), false-positive rate (FPR), and accuracy (ACC), averaging over the simulated point patterns.

Moreover, we compare results obtained fixing $K = \{10,20,30\}$ nearest neighbours, estimating it by means of our proposed procedure, also applying it iteratively up to 3 iterations, in Table \ref{tab:2}.

\begin{table}
\centering
\caption{\label{tab:1} Classification rates averaged over 200 simulated point patterns generated on the unit square with $\mathbb{E}[n_c]$ and $\mathbb{E}[n_f]$ expected number of points for clutter and feature.}
\begin{tabular}{cccc|ccc|ccc|ccc}
 \toprule
&&&&$K$&&iter 1&$K$&&iter 2&$K$&&iter 3\\
\cmidrule(l){5-13}
Scenario&$\mathbb{E}[n_c]$& $\mathbb{E}[n_f]$&Rate& 10 & 20 & 30& 10 & 20 & 30& 10 & 20 & 30 \\ 
  \midrule
Poisson cluster [\ref{scen:1}]& 300& 150& TPR& 0.75& 0.79&0.79 &0.66 &0.65&0.62 &0.53 &0.52 & 0.46\\ 
& & & FPR&  0.53&0.58 &0.61 & 0.42&0.42& 0.39&0.30 & 0.29&0.26\\
& & &ACC&0.56 & 0.54&0.52 &0.61 &0.61& 0.61& 0.64&0.65 &0.65\\
  \midrule
Poisson cluster [\ref{scen:2}] &300& 150& TPR&0.7 &0.75 & 0.75&0.60 &0.61& 0.59&0.47 & 0.46&0.43\\ 
& & & FPR&0.6 & 0.66&0.67 & 0.48&0.48&0.47 &0.35 & 0.33&0.31\\ 
& & &ACC&0.5 &0.48 & 0.47& 0.55&0.54& 0.55& 0.59& 0.60&0.60\\
  \midrule
Poisson [\ref{scen:3}] &300& 150& TPR& 0.93 &0.94& 0.94&0.91 &0.89&0.82 &0.69 & 0.66&0.55\\ 
& & & FPR & 0.34 &0.32&0.32 &0.28 &0.26& 0.23&0.19 &0.17 &0.14\\ 
& & &ACC& 0.75 &0.77&0.77 & 0.78&0.79& 0.79&0.77 & 0.78&0.76\\ 
   \midrule
Poisson [\ref{scen:4}] &300& 20& TPR& 0.98 &0.96  & 0.91  & 1.00  &0.99 & 0.97  & 1.00 &1.00  &0.97 \\ 
& & & FPR&0.58 & 0.42&0.29 &0.63 &0.42& 0.25& 0.64&0.39  &0.22 \\
& & &ACC& 0.46& 0.60 &0.72 &0.40 & 0.60&0.76 &  0.40& 0.63&0.79\\ 
  \bottomrule
\end{tabular}
\end{table}
\begin{table}
\centering
\caption{\label{tab:2} Classification rates averaged over 200 simulated point patterns generated on the unit square with $\mathbb{E}[n_c]$ and $\mathbb{E}[n_f]$ expected number of points for clutter and feature.}
\begin{tabular}{cccc|ccc}
 \toprule
&&&&&$\hat{K}$&\\
\cmidrule(l){5-7}
Scenario&$\mathbb{E}[n_c]$& $\mathbb{E}[n_f]$&Rate& iter 1  &iter 2  &iter 3 \\ 
  \midrule
Poisson cluster [\ref{scen:1}]& 300& 150& TPR&0.80&0.65&0.52\\ 
& & & FPR&  0.65 & 0.42& 0.31\\
& & &ACC& 0.52  &0.60 &0.64\\
  \midrule
Poisson cluster [\ref{scen:2}] &300& 150&TPR& 0.77  &0.63 &0.49\\ 
& & & FPR&  0.69 &0.51 &0.36\\ 
& & &ACC& 0.46  &0.53 &0.59\\
  \midrule
Poisson [\ref{scen:3}] &300& 150& TPR& 0.94& 0.90&  0.70\\ 
& & & FPR& 0.32  &0.27  & 0.18  \\ 
& & &ACC& 0.77 &0.79 & 0.78 \\ 
   \midrule
Poisson [\ref{scen:4}] &300& 20& TPR&    1.00&0.99  &0.98\\ 
& & & FPR& 0.65 &0.44 & 0.27 \\
& & &ACC& 0.39  &0.59  & 0.74\\ 
  \bottomrule
\end{tabular}
\end{table}

Iterating with a fixed $K$ does not improve the classification rates much, while it does with the estimated $\hat{K}$.
In particular, the TPR decreases for the less clustered scenarios (\ref{scen:1}-\ref{scen:3}), indicating that a unique iteration is sufficient in such cases, which is reasonable for these particular cases.
Anyway, the best classification rates are given by the ACC, which indeed increases notably when $\hat{K}$ is estimated compared to when it is fixed. This is true in each considered scenario.
Still discussing increasing iterations, Scenario \ref{scen:4} exhibits the greatest improvement, even when $K$ is fixed. Such improvement is, however, even larger for $\hat{K}$. 

In conclusion, the results on the ACC being in favour of $\hat{K}$, together with the other classification rates being comparable with those of fixed $K$, suggests the usage of the proposed automatic procedure to select the number of nearest neighbours to proceed with the clutter removal procedure.

\section{Case studies}\label{sec:application}

\subsection{Murchison gold data}
\begin{figure}[H]
\centering
\includegraphics[width=.45\textwidth]{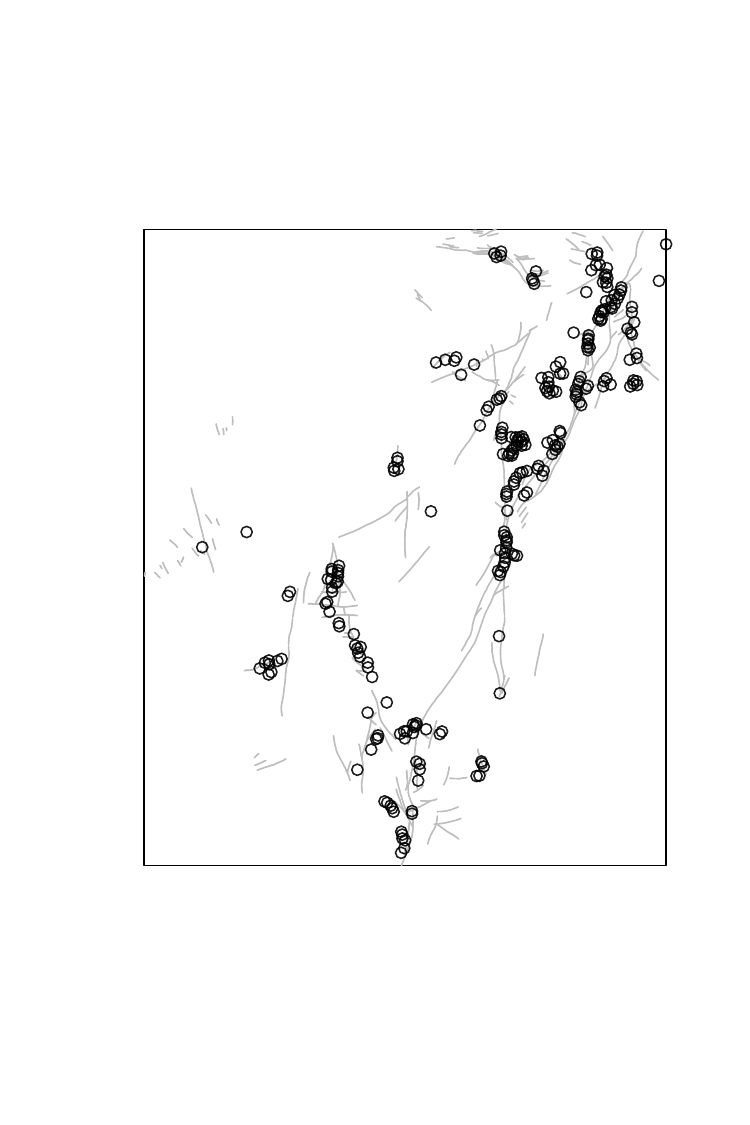}
	\caption{Murchison gold data: in grey, the locations of geological faults, and in black, the locations of gold deposits.}
	\label{fig:stopping4}
\end{figure}

The Murchison geological survey data shown in Figure \ref{fig:stopping4} record the spatial locations of gold deposits and associated geological features in the Murchison area of Western Australia.

They are extracted from a regional survey (scale 1:500,000) of the Murchison area carried out by the Geological Survey of Western Australia \citep{watkins1990geological}. The point pattern recorded is the known locations of gold deposits, and they come with the known or inferred locations of geological faults. The study region is contained in a $330 \times 400$ kilometer rectangle. At this scale, gold deposits are point-like, i.e. their spatial extent is negligible. Gold deposits are strongly associated with greenstone bedrock and faults, but the geology is three-dimensional, and the survey data are a two-dimensional projection. The survey may not have detected all existing faults because they are usually not observed
directly; they are observed in magnetic field surveys or geologically inferred from discontinuities in
the rock sequence. These data were analysed in \cite{foxall2002nonparametric,brown2002bivariate} and \cite{groves2000late,knox1997gold}. The main aim is usually to predict the
intensity of the point pattern of gold deposits from the more easily observable fault pattern.
We apply the EM procedure iteratively, which stops at the second iteration thanks to the proposed stopping criterion.
Note that the nearest neighbours selected at each iteration are 26 and 7.

The points classified as features clearly identify an underlying fault.

 \begin{figure}[H]
\centering
\includegraphics[width=\textwidth]{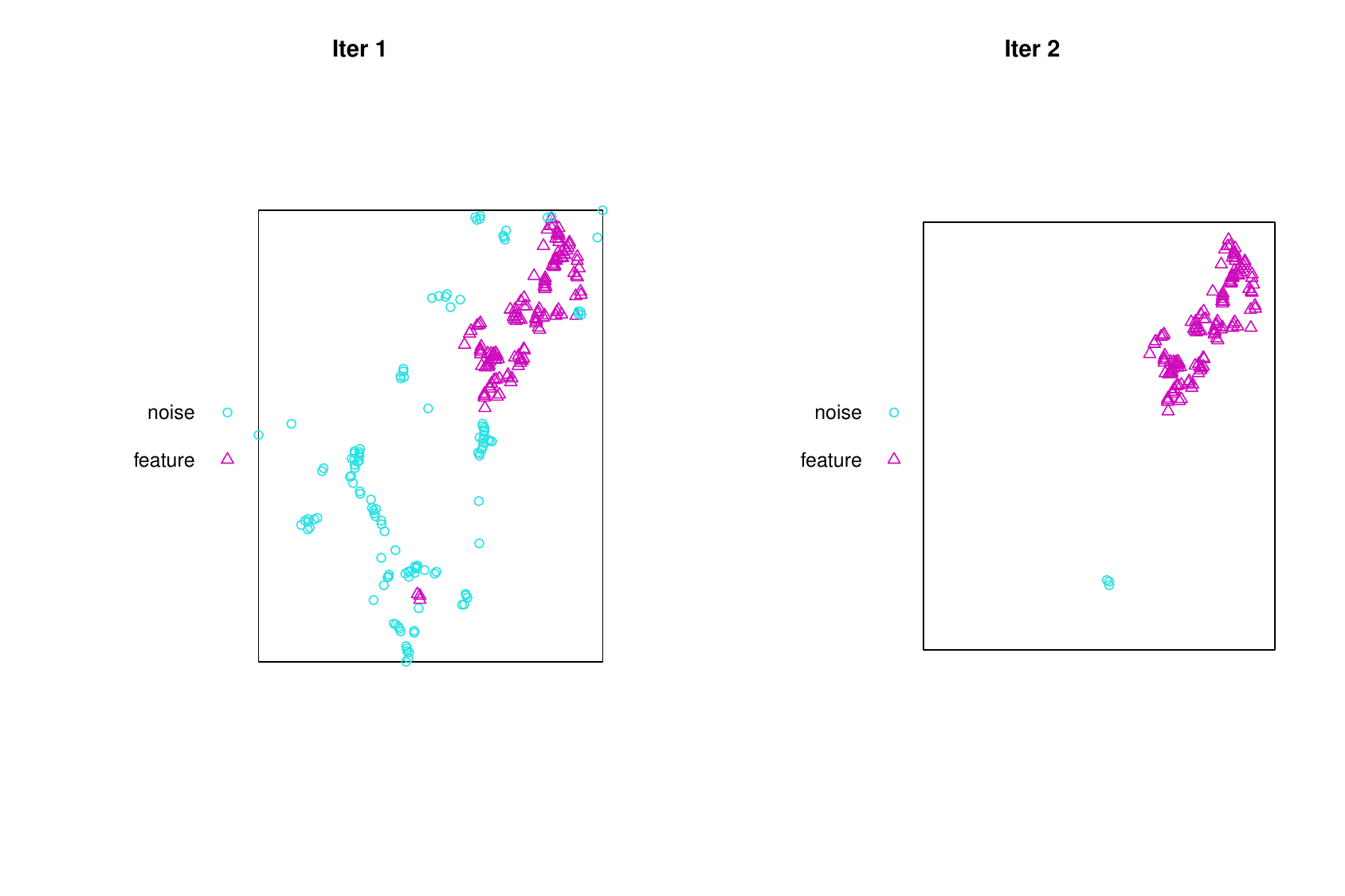}
	\caption{Output of the proposed iterative procedure up to 2 iterations. Blue denotes the \textit{clutter}/\textit{noise} points, and pink denotes the \textit{feature} points.}
	\label{fig:stopping3}
\end{figure}

\subsection{Detecting seismic faults}
\cite{dasgupta1998detecting} considered the problem of detecting seismic faults based on an earthquake catalogue. The idea is that earthquake epicentres occur along seismically active faults and are measured with some error. So, over time, observed earthquake epicentres should be clustered along such faults. \cite{dasgupta1998detecting} considered an earthquake catalogue recorded over a 40,000  km2 region of the central coast ranges in California from  1962-1981 \citep{mckenzie1982bulletin}. An advantage of looking at this region is that the known fault structure is well documented.
 \cite{dasgupta1998detecting}  selected a classification  with seven clusters (six non-noise clusters and one noise
 cluster) because the BIC attains a local maximum there and the successive differences in the BIC values are small thereafter.
They found that the classification obtained using six (non-noise) clusters corresponds well with the available documentation of faults in the region of interest. One or two clusters do not correspond to any of the documented faults. 

An application of 5th NN clutter removal produced the results on \cite{Byers1998}. One key difference is the isolated cluster in the bottom right that NN methods pick up but that the connected component part of Allard and Fraley's method leaves out. This cluster is treated as one end of a linear cluster of earthquakes in the analysis of \cite{dasgupta1998detecting}.
 They end up filling in the sparse part between it and other clusters with clutter to produce the linear form that they search for. It would seem that the MClust-EM method is  more suited to finding features such as faults that are supposed to be roughly linear, but the differences exposed here  show that less-structured methods do have contributions to
 make in structured situations.

\begin{figure}[H]
\centering
\vspace{-2.5cm}
\includegraphics[width=.5\textwidth]{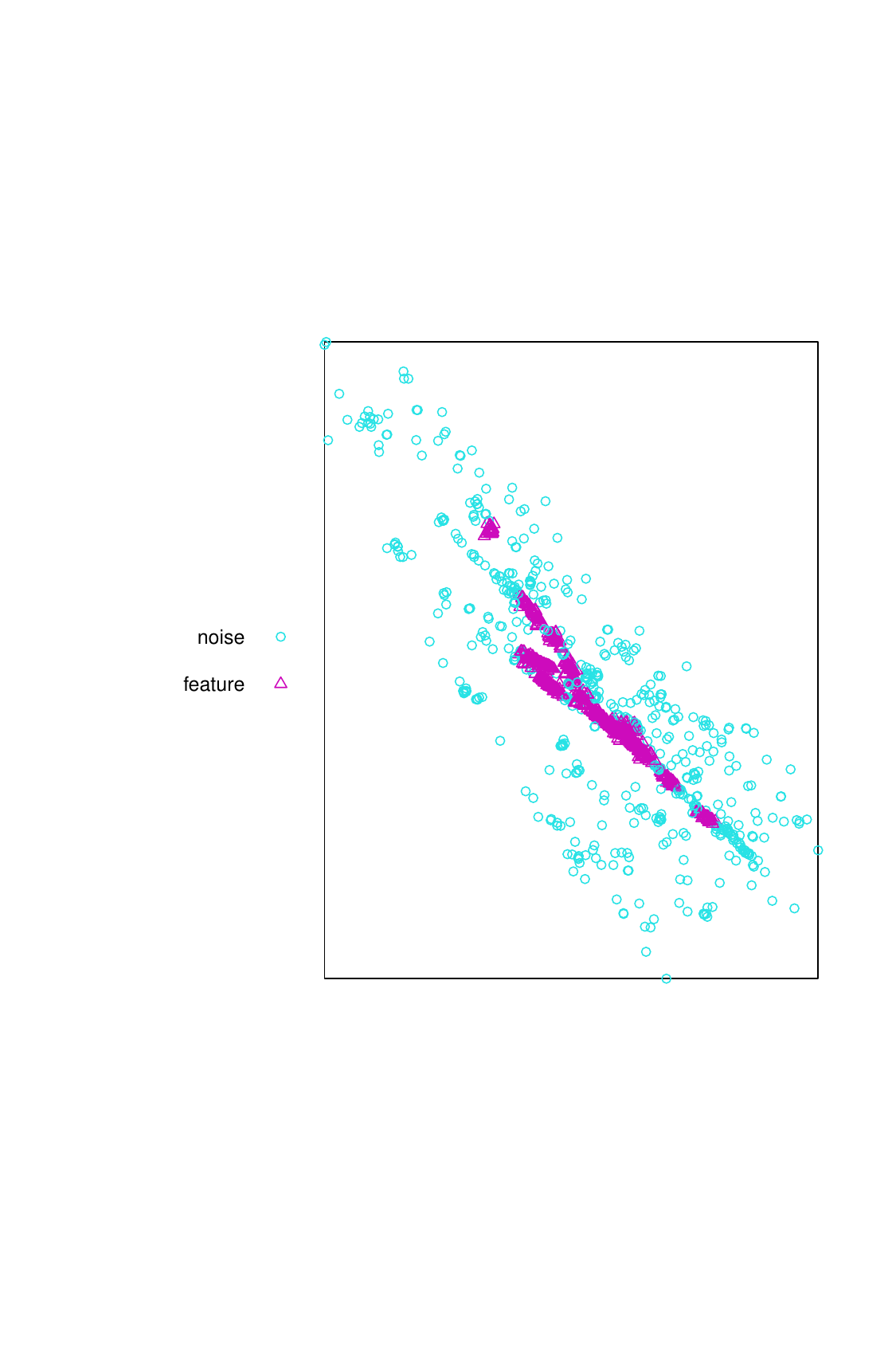}
\vspace{-2.5cm}
	\caption{Output of the proposed iterative procedure applied to the analysed earthquake data. Blue denotes the \textit{clutter}/\textit{noise} points, and pink denotes the \textit{feature} points.}
	\label{fig:step_appl}
\end{figure}

We analyse the same catalogue of North California earthquakes of magnitude at least 2.5, available from \url{https://ncedc.org/ncedc/catalog-search.html}.
We proceed to run the proposed iterative procedure, which stops at the first one.
The nearest neighbour selected is 19. Figure \ref{fig:step_appl} displays the detected feature points, indicating the major underlying San Andreas Fault.
 
\section{Conclusions}\label{sec:conclusions}

In this paper, we have addressed the problem of selecting the $K$th nearest neighbour in the clutter removal procedure for spatial point processes, as well as the problem of finding a suitable stopping criterion when applying the algorithm iteratively to get better results.

The methods proposed in this paper build upon the existing classification method of \cite{Byers1998}, which models the Kth nearest neighbour distances of an observed point pattern made by the superimposition of clutter and feature points by means of a mixture distribution. 

The contributions of this paper are twofold. 
Firstly, we introduced an automated method for determining the optimal number of nearest neighbours, utilizing segmented regression models. This enhancement aimed to formalize such selection and to refine the classification process overall, making it completely automatic and, therefore, reproducible. 
Secondly, with the aim in mind of improving the results in terms of classification, we explore the context of iteratively applying the classification procedure. 
We do so by introducing a stopping criterion that minimises the overall entropy measure of cluster separation between clutter and feature points at each iteration and stops whenever we get no further improvement in such sense.

Through simulations and real-world case studies involving environmental data, we demonstrated the efficacy of our proposed procedures, showcasing their utility in practical applications.
Performing similarly to the benchmark methodology in terms of accuracy, our proposed selection method represents a convenient automatic procedure to apply in real data applications when the best number of nearest neighbours to consider is unknown.
These enhancements not only provide more accurate feature detection but also offer a systematic and automated approach for refining the classification process, thereby enhancing the overall reliability and applicability of the method in various spatial contexts.

Note that these methodological improvements are applicable to all those scenarios where features are superimposed on clutter and, therefore, modelled as two overlapping Poisson processes, including the context of point processes linear networks and that of spatio-temporal point processes. 

For this reason, future works will adapt the proposed procedure to such a more complex context of point patterns.

Finally, another promising extension worth investigating in the future is the alteration of the EM algorithm to search for $r > 2$ groups, each with a different rate.
\cite{Byers1998} states that such a scenario's performance is in line with that of the two-rate case. 
This could be useful when each group might correspond to a set of features with a different density, e.g. seismic faults with different earthquake frequencies.

 \section*{Fundings}
This work has been supported by the Targeted Research Funds 2023 (FFR 2023) of the University of Palermo (Italy) and by the PNRR project, grant agreement No PE0000018 - “GRINS - Growing Resilient, INclusive and Sustainable”.

\bibliography{example}

\begin{thebibliography}{}

\bibitem[Allard and Fraley, 1997]{1997-AF}
Allard, D. and Fraley, C. (1997).
\newblock Nonparametric maximum likelihood estimation of features in spatial
  point processes using vorono$\ddot{\mbox{i}}$ tessellation.
\newblock {\em Journal of the American Statistical Association},
  92(440):1485--1493.

\bibitem[Banfield and Raftery, 1993]{banfield1993model}
Banfield, J.~D. and Raftery, A.~E. (1993).
\newblock Model-based gaussian and non-gaussian clustering.
\newblock {\em Biometrics}, pages 803--821.

\bibitem[Brown et~al., 2002]{brown2002bivariate}
Brown, W., Gedeon, T., Baddeley, A., and Groves, D. (2002).
\newblock Bivariate j-function and other graphical statistical methods help
  select the best predictor variables as inputs for a neural network method of
  mineral prospectivity mapping.
\newblock In {\em Bivariate J-function and other graphical statistical methods
  help select the best predictor variables as inputs for a neural network
  method of mineral prospectivity mapping}, pages 257--268. International
  Association for Mathematical Geology.

\bibitem[Byers and Raftery, 1998]{Byers1998}
Byers, S. and Raftery, A.~E. (1998).
\newblock Nearest-neighbor clutter removal for estimating features in spatial
  point processes.
\newblock {\em Journal of the American Statistical Association},
  93(442):577--584.

\bibitem[Celeux and Soromenho, 1996]{celeux1996entropy}
Celeux, G. and Soromenho, G. (1996).
\newblock An entropy criterion for assessing the number of clusters in a
  mixture model.
\newblock {\em Journal of classification}, 13(2):195--212.

\bibitem[Cox and Isham, 1980]{cox1980point}
Cox, D.~R. and Isham, V. (1980).
\newblock {\em Point processes}, volume~12.
\newblock CRC Press.

\bibitem[Cressie, 2015]{cressie2015statistics}
Cressie, N. (2015).
\newblock {\em Statistics for spatial data}.
\newblock John Wiley \& Sons.

\bibitem[Daley et~al., 2003]{daley2003introduction}
Daley, D.~J., Vere-Jones, D., et~al. (2003).
\newblock {\em An introduction to the theory of point processes: volume I:
  elementary theory and methods}.
\newblock Springer.

\bibitem[Dasgupta and Raftery, 1998]{dasgupta1998detecting}
Dasgupta, A. and Raftery, A.~E. (1998).
\newblock Detecting features in spatial point processes with clutter via
  model-based clustering.
\newblock {\em Journal of the American statistical Association},
  93(441):294--302.

\bibitem[Dempster et~al., 1977]{DLR1977}
Dempster, A.~P., Laird, N.~M., and Rubin, D.~B. (1977).
\newblock Maximum likelihood from incomplete data via the em algorithm.
\newblock {\em Journal of the Royal Statistical Society. Series B
  (Methodological)}, 39(1):1--38.

\bibitem[Diggle et~al., 1976]{diggle1976statistical}
Diggle, P.~J., Besag, J., and Gleaves, J.~T. (1976).
\newblock Statistical analysis of spatial point patterns by means of distance
  methods.
\newblock {\em Biometrics}, pages 659--667.

\bibitem[Foxall and Baddeley, 2002]{foxall2002nonparametric}
Foxall, R. and Baddeley, A. (2002).
\newblock Nonparametric measures of association between a spatial point process
  and a random set, with geological applications.
\newblock {\em Journal of the Royal Statistical Society: Series C (Applied
  Statistics)}, 51(2):165--182.

\bibitem[Gonz{\'a}lez et~al., 2021]{gonzalez2021classification}
Gonz{\'a}lez, J.~A., Rodr{\'\i}guez-Cort{\'e}s, F.~J., Romano, E., and Mateu,
  J. (2021).
\newblock Classification of events using local pair correlation functions for
  spatial point patterns.
\newblock {\em Journal of Agricultural, Biological and Environmental
  Statistics}, 26(4):538--559.

\bibitem[Groves et~al., 2000]{groves2000late}
Groves, D.~I., Goldfarb, R.~J., Knox-Robinson, C.~M., Ojala, J., Gardoll, S.,
  Yun, G.~Y., and Holyland, P. (2000).
\newblock Late-kinematic timing of orogenic gold deposits and significance for
  computer-based exploration techniques with emphasis on the yilgarn block,
  western australia.
\newblock {\em Ore Geology Reviews}, 17(1-2):1--38.

\bibitem[Illian et~al., 2008]{illian:penttinen:stoyan:stoyan:08}
Illian, J., Penttinen, A., Stoyan, H., and Stoyan, D. (2008).
\newblock {\em Statistical Analysis and Modelling of Spatial Point Patterns},
  volume~70.
\newblock John Wiley \& Sons.

\bibitem[Knox-Robinson and Groves, 1997]{knox1997gold}
Knox-Robinson, C. and Groves, D. (1997).
\newblock Gold prospectivity mapping using a geographic information system
  (gis) with examples from the yilgarn block of western australia.
\newblock {\em Chronique de la Recherche Mini{\`e}re}, (529):127--138.

\bibitem[McKenzie et~al., 1982]{mckenzie1982bulletin}
McKenzie, M., Miller, R., and Uhrhammer, R. (1982).
\newblock Bulletin of the seismographic stations.
\newblock {\em University of California, Berkeley}, 53(1-2).

\bibitem[Moller and Waagepetersen, 2003]{moller2003statistical}
Moller, J. and Waagepetersen, R.~P. (2003).
\newblock {\em Statistical inference and simulation for spatial point
  processes}.
\newblock CRC press.

\bibitem[Muggeo, 2003]{muggeo2003estimating}
Muggeo, V. M.~R. (2003).
\newblock Estimating regression models with unknown break-points.
\newblock {\em Statistics in Medicine}, 22(19):3055--3071.

\bibitem[Muggeo, 2008]{muggeo2008segmented}
Muggeo, V. M.~R. (2008).
\newblock segmented: An {R} package to fit regression models with broken-line
  relationships.
\newblock {\em R news}, 8(1):20--25.

\bibitem[{R Core Team}, 2023]{R}
{R Core Team} (2023).
\newblock {\em R: A Language and Environment for Statistical Computing}.
\newblock R Foundation for Statistical Computing, Vienna, Austria.

\bibitem[Ripley, 2005]{ripley2005spatial}
Ripley, B.~D. (2005).
\newblock {\em Spatial statistics}.
\newblock John Wiley \& Sons.

\bibitem[Schoenberg and Tranbarger, 2008]{schoenberg2008description}
Schoenberg, F.~P. and Tranbarger, K.~E. (2008).
\newblock Description of earthquake aftershock sequences using prototype point
  patterns.
\newblock {\em Environmetrics: The official journal of the International
  Environmetrics Society}, 19(3):271--286.

\bibitem[Tranbarger~Freier and Schoenberg, 2010]{tranbarger2010computation}
Tranbarger~Freier, K.~E. and Schoenberg, F.~P. (2010).
\newblock On the computation and application of prototype point patterns.
\newblock {\em The Open Applied Informatics Journal}, 4(1).

\bibitem[Watkins and Hickman, 1990]{watkins1990geological}
Watkins, K.~P. and Hickman, A.~H. (1990).
\newblock {\em Geological evolution and mineralization of the Murchison
  Province, Western Australia}, volume~1.
\newblock Geological Survey of Western Australia.

\end{thebibliography}

\end{document}